\documentstyle[preprint,aps,floats,epsfig]{revtex}
\tightenlines
\begin{document}

\title{Low-mass dileptons and dropping rho meson mass}
\author{E. L. Bratkovskaya$^1$ and C. M. Ko$^2$}
\address{$^1$Institute f\"ur Theoretiche Physik, Universit\"at
Giessen, 35392 Giessen, Germany}
\address{$^2$Cyclotron Institute and Physics Department, \\
Texas A\&M University, College Station, Texas 77843}
\maketitle

\begin{abstract}
Using the transport model, we have studied dilepton production from
heavy-ion collisions at Bevalac energies. It is found that the
enhanced production of low-mass dileptons observed in the
experiment by the DLS collaboration cannot be explained by the
dropping of hadron masses, in particular the $\rho$-meson mass, in
dense matter.

\end{abstract}

\vspace{0.5cm}
\vspace{0.5cm}


One of the most exciting physics one expects to learn from
relativistic heavy-ion collisions is the possibility to create
nuclear matter at densities larger than that in the center of
normal nuclei and possibly comparable to that in the interior of
neutron stars \cite{sg86,giessen,koli96}. This thus offers the
opportunity to study if the nuclear equation of state becomes
softened at high densities as required from studies of supernova
explosions. The high density nuclear matter created in these
collisions also makes it possible to study how spontaneously broken
chiral symmetry, which is characterized by a large value of the
quark condensate in the vacuum, is partially restored when the
temperature and density of matter becomes high, as predicted by
theoretical studies \cite{condensate}. Unfortunately, a direct
relation of the quark condensate to physical observables has not
been rigorously established \cite{kkl97}. One conjecture has been
suggested by Brown and Rho \cite{br91} that masses of non-strange
hadrons such as nucleon, $\rho$, and $\omega$, are reduced in the
nuclear medium and are proportional to the in-medium quark
condensate. Dropping hadron in-medium masses have also been found
in some theoretical models, such as the QCD sum rules
\cite{hatsuda}, the quark-meson coupling model \cite{saito94}, and
the hadronic model including vacuum polarization effects
\cite{jean94}. Such scaled in-medium hadron masses have been shown
to lead to significantly improved understanding of many nuclear
phenomena, such as the enhanced axial charge transitions seen in
heavy nuclei \cite{war92} and the quenching in the longitudinal
response function of nuclei \cite{ste97}.

A more direct evidence for a dropping $\rho$-meson mass seems to be
provided by the enhanced production of dileptons with invariant
mass around 400-500 MeV above known sources of dileptons in the
CERES experiment of heavy-ion collisions at CERN SPS \cite{aga95}.
In theoretical studies \cite{li95,cass95,sriv96,koch96,brat97}
based on various models ranging from a simple thermal model to
sophisticated transport models, dilepton production has been
investigated by including contributions not only from the Dalitz
decay of mesons and direct dilepton decay of vector mesons but also from 
pion-pion annihilation. The latter is practically absent in
proton-nucleus collisions and is found to account for only about
half of the observed enhancement in heavy-ion collisions. Including
contributions from other processes such as the $a_1$ decay
\cite{haglin96} and pion-rho scattering \cite{baier97} does not
help much. On the other hand, allowing for the reduction of the
$\rho$ meson in-medium mass in the relativistic transport model
\cite{li95,cass95,brat97}, it has been shown that it can indeed
lead to a further enhancement of low-mass dileptons as observed in
the experimental data. On the other hand, it has also been claimed
\cite{rapp96} that if the $\rho$ meson -- which in free space has a
broad decay width to two pions -- melts in the medium due to
additional decay channels such as the resonance-hole states
\cite{rapp97,klin97,pete98}, would also significantly increase the
yield of low-mass dileptons. However, in the dynamical studies of
Ref. \cite{cass98} the $\rho$ spectral function, which consists of
effects due to both medium modifications of pions and the
baryon-resonance-nucleon-hole excitations, is introduced through
pion-pion annihilation to dileptons via the vector meson dominance.
Because of the large pion to nucleon ratio (about 5) in heavy-ion
collisions at SPS energies, the effect of resonance-hole
contributions to dilepton production is thus overestimated in these
studies.

Dileptons have also been measured in heavy-ion collisions at the
Bevalac by the DLS collaboration \cite{dls1,dls2} at incident
energies that are two orders-of-magnitude lower than that at SPS.
Although the first published data \cite{dls1} based on a limited
data set are consistent with the results from transport model
calculations \cite{xiong,wolf} that include $pn$ bremsstrahlung,
$\pi^0$, $\eta$ and $\Delta$ Dalitz decay and pion-pion
annihilation, a recent analysis \cite{dls2}, including the full
data set, shows a considerable increase in the cross section, which
is now more than a factor of five above these theoretical
predictions even after including also contributions from the decay
of $\rho$ and $\omega$ that are produced directly from
nucleon-nucleon and pion-nucleon scattering in the early reaction
phase \cite{brat,Ernst}. With an in-medium rho spectral function as
that used in Ref. \cite{cass98} for dilepton production from
heavy-ion collisions at SPS energies, dileptons from the decay of
both directly produced $\rho$'s and pion-pion annihilation have
been considered, and a factor of two enhancement has been obtained
compared to the case of using a free $\rho$-spectral function.
Since the pion to nucleon ratio in heavy-ion collisions at Bevalac
energies is much smaller than one (about 0.2), contrary to that in
heavy-ion collisions at SPS energies, the contribution of pion-pion
annihilation instead of baryon resonance Dalitz decay is
overestimated in the spectral function analysis in Ref.
\cite{brat}. As dropping hadron masses in dense matter have been
seen to account for the observed enhancement of low-mass dileptons
in heavy-ion collisions at SPS energies, it is of interest to see
if this can also explain the enhanced production of dileptons at
Bevalac energies.

We shall base our study on the Hadron String Dynamics (HSD) model
\cite{Ehehalt}, which describes the collision dynamics by
propagating nucleons in a mean-field potential given by the
attractive scalar and repulsive vector potentials. Free
nucleon-nucleon cross sections are used in treating their
collisions, which include both elastic and inelastic ones. The
latter leads to the production of both meson resonances such as
$\rho$ and $\omega$ and baryon resonances such as $\Delta(1232)$,
$N^*$(1440), and $N^*(1535)$. The decay of baryon resonances then
leads to the production of pions and $\eta$'s. Scattering of pions
with nucleons is also included, leading to the production of meson
and baryon resonances as well. This model has been shown to give
very good descriptions of the measured yield of pions and $\eta$'s
as well as the rapidity and transverse momentum distributions of
hadrons \cite{casspr}.

To study dilepton production, we include contributions from $pn$
and $\pi N$ bremsstrahlung; Dalitz decay of $\pi^0$, $\eta$,
$\omega$, $\Delta$, and $N^*(1440)$; and direct decays of $\rho$
and $\omega$. All contributions will be treated in the standard way
as in Refs. \cite{xiong,wolf,brat,liko}. However, since vector
meson production in nucleus-nucleus collisions at Bevalac energies,
which are below the production subthreshold in nucleon-nucleon
collisions, is dominated by the $\pi N$ channels \cite{Brat97mt},
we shall pay a special attention to this contribution. In
particular, it has been recently shown that the rho meson couples
strongly to $N^*(1520)$ \cite{pete98,brown}, we shall thus include
additionally this effect in rho meson production from $\pi N$
scattering, as it has been overlooked in previous work.

Specifically, the isospin-averaged production cross section of
a neutral $\rho$ meson from $\pi N$ scattering through $N^*(1520)$
at a center-of-mass energy $\sqrt{s}$ is
\begin{eqnarray}\label{pnrn}
{d\sigma_{\pi N\to\rho N}(s,M)\over dM}={\rm S_{\pi N}} {4\pi \over
k_\pi^2}{s \Gamma_{N^*\to \pi N}(\sqrt{s})\over (s-m_{N^*}^2)^2+
s{\Gamma_{\rm tot}^{N^*}}^2(\sqrt{s})} {d\Gamma_{N^*\to\rho
N}(\sqrt{s},M)\over dM}\label{piNrhoM},
\end{eqnarray}
with
\begin{eqnarray}
{\rm S_{\pi N}}={1\over 3} {(2J_{N^*}+1) \over (2J_\pi +1) (2J_N +1)}
{(2I_{N^*}+1) \over (2I_\pi +1) (2I_N +1)},\nonumber
\end{eqnarray}
where $J_{N^*}=3/2, J_\pi=0, J_N=1/2, I_{N^*}=1/2, I_\pi=1$, and
$I_N=1/2$. In Eq. (\ref{pnrn}), $k_\pi$ is the pion three-momentum
in the center of mass of pion and nucleon.

The partial decay width of a $N^*(1520)$ of mass $\sqrt{s}$ to a
nucleon and a $\rho$ meson of mass $M$ is given in Ref.
\cite{pete98}, i.e.,
\begin{eqnarray}\label{nsrn1}
{d\Gamma_{N^*\to\rho N}(\sqrt{s},M)\over dM} =
\left({f_{N^*N\rho}\over m_\rho}\right)^2 \ {1\over \pi} \
M {m_N\over \sqrt{s}} \ k_\rho
(2\omega_\rho^2+M^2) A(M)\ F(k_\rho^2),
\end{eqnarray}
where $f_{N^*N\rho}\sim 7$ is the coupling constant, $k_\rho$
denotes the three-momentum of the $\rho$ meson in the rest frame of
$N^*(1520)$, and $\omega_\rho^2=M^2+k_\rho^2$ is its energy. The
$\rho$ spectral function is denoted by $A(M)$ and has a
Breit-Wigner form,
\begin{eqnarray}\label{breit}
A(M) = {1\over \pi} {m_\rho \Gamma_{\rm tot}^\rho(M) \over
(M^2-m_\rho^2)^2 + (m_\rho {\Gamma_{\rm tot}^{\rho}(M)})^2}.
\label{BW}\end{eqnarray}
The form factor $F(k_\rho^2)$ is taken to have a monopole form
\begin{eqnarray}
 F(k_\rho^2) ={\Lambda^2\over \Lambda^2+k_\rho^2},
\label{fff}\end{eqnarray}
with $\Lambda=1.5 \ {\rm GeV}$.

The $\rho$ meson total width $\Gamma_{\rm tot}^\rho(M)$ is given by
the sum of the width due to decay to two pions and that due to
absorption by nucleons and elastic scattering, i.e.,
\begin{eqnarray}
\Gamma_{\rm tot}^\rho(M) &=& \Gamma_{\rho\to \pi\pi}(m_\rho) \
\left(m_\rho\over M\right) \ \left(k_\pi(M)\over k_\pi(m_\rho)\right)^3
\label{rhowidth}\\
&+&\sigma_{\rho N}(s,M) v \gamma \rho_N,
\nonumber\end{eqnarray}
where $\Gamma_{\rho\to \pi\pi}(m_\rho)\sim 0.15$ GeV and $k_\pi$ is
the pion momentum in the rest frame of the $\rho$ meson. The second
term corresponds to the collisional broadening width and is given
by the product of $\rho$-nucleon total cross section $\sigma_{\rho
N}(s,M)$, the $\rho$ meson velocity $v$ in the rest frame of the
nucleon current, the associated Lorentz factor $\gamma$, and the
nucleon density $\rho_N$. Based on the resonance model, the
$\rho$-nucleon total cross section has been determined in Ref.
\cite{Kondr98}, and the result can be parameterized as
\begin{equation}\label{rhon}
\sigma_{\rho N}(s,M)=26 + 15 k_\rho^{-1.2}~[\rm mb],
\end{equation}
with $k_\rho$ [GeV/c] the rho meson three-momentum in the rest
frame of the nucleon current. At normal nuclear matter density,
this gives an average rho meson collisional width of about 75~MeV
whereas in central $^{40}$Ca+$^{40}$Ca reactions at 1 A$\cdot$GeV
the average collisional width amounts to $\simeq 140$~MeV.

The partial decay width of a $N^*(1520)$ to $\pi N$ is \cite{pete98}
\begin{eqnarray}\label{partial}
\Gamma_{N^*\to \pi N} (\sqrt{s}) = \Gamma_0
\left(k_\pi(\sqrt{s})\over k_\pi(m_{N^*})\right)^{2l+1},
\label{partW}
\end{eqnarray}
with $k_\pi(m)$ the pion three-momentum in the rest frame of
$N^*(1520)$, $l=2$, and $\Gamma_0=0.095$ GeV.

The total decay width of the $N^*(1520)$ is given by the sum of the
partial decay widths to pion and $\rho$, i.e., $\Gamma_{\rm
tot}^{N^*} =
\Gamma_{N^*\to \pi N} + \Gamma_{N^*\to \rho N}$, where
$\Gamma_{N^*\to \rho N}$ is obtained from
\begin{eqnarray}\label{nsrn2}
\Gamma_{N^*\to \rho N} (\sqrt{s}) = \int\limits_{2m_\pi}^{\sqrt{s}-m_N}
dM{d\Gamma_{N^*\to\rho N}(\sqrt{s},M)\over dM}.
\label{IntGrho}\end{eqnarray}

The dilepton production cross section from the direct decay of the
rho meson produced from $\pi N$ scattering through $N^*(1520)$ is
then
\begin{eqnarray}
{d\sigma_{\pi N\to e^+e^-N}(s,M)\over dM}=
{d\sigma_{\pi N\to\rho N}(s,M)\over dM}
{\Gamma_{\rho\to e^+e^-}(M) \over \Gamma_{\rm tot}^\rho(M)}.
\label{piNdil}
\end{eqnarray}
In the above, $\Gamma_\rho^{e^+e^-}(M)$ is the dilepton decay width
of a neutral rho meson of mass $M$. From the vector dominance
model, it is given by
\begin{equation}\label{ree}
\Gamma_\rho^{e^+e^-}(M)=8.8\times 10^{-6}\frac{m_\rho^4}{M^3}.
\end{equation}

\begin{figure}
\epsfig{file=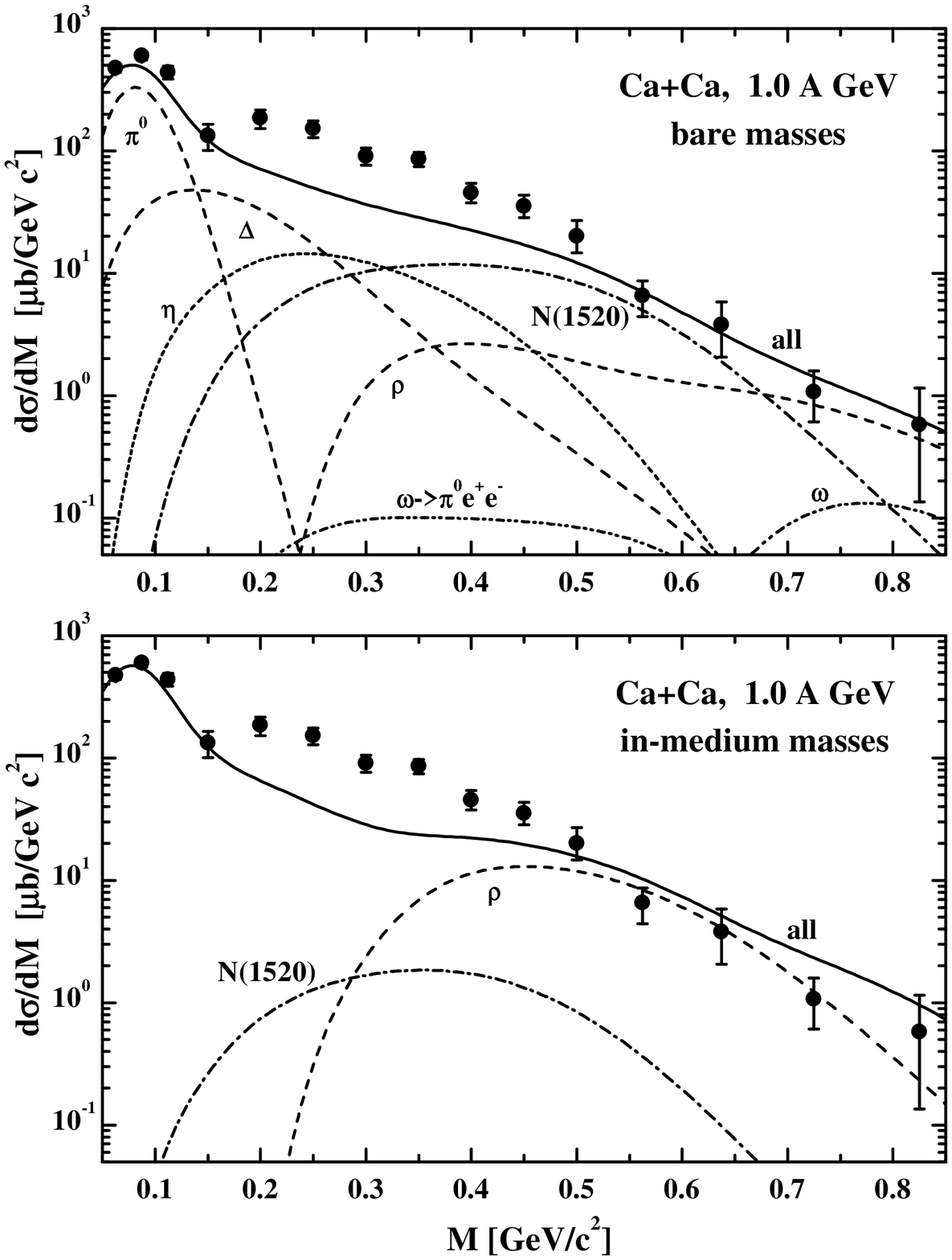,width=15cm}
\caption{The dilepton invariant mass spectrum from $^{40}$Ca+$^{40}$Ca
collisions at 1 A$\cdot$GeV with free (upper part) and in-medium
(lower part) hadron masses in comparison to the DLS data
\protect\cite{dls2}.}
\label{cadil}
\end{figure}

For the case of free hadron masses, the dilepton invariant mass
spectrum from $^{40}$Ca+$^{40}$Ca collisions at 1 A$\cdot$GeV and
after correcting for the experimental acceptance filter (version
4.1) is shown in the upper part of Fig. \ref{cadil}. It is seen
that with increasing dilepton invariant mass, the dominating
contribution shifts from $\pi^0$ Dalitz decay to $\Delta$ Dalitz
decay, to $\eta$ Dalitz decay, and finally to direct $\rho$ decay,
as in Refs. \cite{brat,Ernst}. In particular, the contribution from
the direct decay of rho mesons produced from $\pi N$ scattering
through the $N^*(1520)$ (dot-dashed curve) is most important in the
mass region $0.35<M<0.75$ GeV/c$^2$ and exceeds that from other
$\rho$ production channels (dashed curve) -- $\pi\pi$ annihilation,
pion-baryon (without $N^*(1520)$) and baryon-baryon collisions.
This is different from heavy-ion collisions at CERN SPS, where
dilepton production from the $\pi\pi\to \rho$ channel is more
important than that from the $\pi N \to \rho N$ channel as a result
of the large pion to nucleon ratio in these collisions. Compared
with the experimental data the theoretical results for the total
dilepton spectrum are, however, about a factor of three lower in
the invariant mass region $0.2<M<0.5$ GeV/c$^2$ and practically the
same as in the spectral function approach of Ref. \cite{brat}.

We note that contributions of other baryon resonances, e.g., the
$N^*(1700)$ which also couples strongly to the rho meson
\cite{friman}, to low-mass dileptons are negligibly small as a
result of their large masses and/or relatively weak coupling to the
rho meson and nucleon.

To see how the results are modified by medium effects, we introduce
the rho/omega meson in-medium masses as in Ref. \cite{li95,cass95}
but keep the baryon masses unchanged,
\begin{eqnarray}
m_{\rho/\omega}^* &\sim& m_{\rho/\omega}(1-0.18\rho_B/\rho_0).
\end{eqnarray}
According to Ref. \cite{brown}, part of the decrease of the $\rho$
meson mass in nuclear medium can be accounted for by the attractive
interaction due to $N^*(1520)$-particle-nucleon-hole polarization.
The contribution to dilepton production from $\pi N$ scattering is
then computed from Eqs. (\ref{pnrn})-(\ref{ree}) using in-medium
rho meson mass. Specifically, the rho meson mass in Eqs.
(\ref{breit}) and (\ref{ree}) are replaced by the in-medium one.
Also, $k_\rho$ in Eqs. (\ref{nsrn1}) and (\ref{rhon}) are evaluated
with the in-medium mass.

The reduction of omega mass enhances its production but does not
significantly increase the yield of low mass dileptons due to the
small partial decay width of omega meson to dileptons compared to
that of rho meson. Omega mesons thus contribute to dilepton
production mainly at freeze out when they have regained their free
mass. Furthermore, omegas can be absorbed by nucleons. From omega
photoproduction data, analyses based on the Vector-Dominance model
or the additive quark model give an omega-nucleon total cross
section of $25\sim 30$ mb at omega momenta above 1 GeV/c, similar to
that for the rho-nucleon total cross section obtained in the same
model \cite{bauer78}. Also, for omega at finite momenta, its
absorption cross section through the reaction $\omega N\to \pi N$
can be obtained from the inverse reaction $\pi N\to\omega N$, which
has an empirical value similar to that for $\pi N\to\rho N$ (see
e.g. \cite{sib96}). On the other hand, for omega-nucleon scattering
near threshold, Friman \cite{Friman_omN} has found from the $\pi
N\to\omega N$ data that the imaginary part of the omega-nucleon
scattering length is about a factor of seven smaller than that for the
rho meson, giving thus a much smaller omega absorption cross
section than for the rho meson. The latter result is, however,
expected to change appreciably if one also takes into account final
states that consists of more than one pion \cite{cassing}. Since
the omega contribution to low-mass dileptons is unimportant in
heavy ion collisions at Bevalac energies, we have thus adopted
the simple assumption that its interaction cross section with a
nucleon is similar to that for a rho meson, i.e., Eq.~(\ref{rhon}).
We then find that the contribution of omega mesons to dileptons are
not much affected by the change of their masses and thus remains
unimportant.

The dilepton invariant mass spectrum from the same reaction for the
case of dropping rho meson mass is shown in the lower part of Fig.
\ref{cadil}. We find that with dropping rho meson mass low-mass
dilepton production from $\pi N$ scattering through $N^*(1520)$ is
substantially reduced due to a significant increase of its width.
Although dileptons from other $\rho$ production channels are
enhanced, they are not sufficient to compensate for the reduction
due to the broadening of N(1520). Including also the contributions from 
the Dalitz decays of $\pi^0$, $\Delta$, $\eta$, and $\omega$
as well as the direct decay of $\omega$, which are not much
affected by the reduction of hadron masses, the total theoretical
dilepton spectrum (solid curve) remains about a factor of three
below the experimental data for $0.2\le M \le 0.5$~GeV. Similar
results are obtained if we also allow the nucleon and N(1520)
masses to decrease according to the constituent quark model, i.e.,
a factor 3/2 reduction relative to that for the $\rho$ meson.

In conclusion, we have shown using the HSD transport model that
although dilepton production from pion-nucleon scattering through
the $N^*(1520)$ is important in heavy-ion collisions at Bevalac
energies, as the nucleon is the most abundant particles, including
its contribution cannot explain the observed enhancement by the DLS
collaboration. Allowing the reduction of rho meson mass in dense
matter enhances dilepton production from other rho production
channels, but it reduces significantly that from $\pi N\to
N^*(1520)\to\rho N$ as a result of the broadening of $N^*(1520)$
width, leading to a total dilepton spectrum similar to that without
dropping hadron masses. Thus, unlike the enhanced low mass
dileptons observed in heavy ion collisions at SPS energies, which
is dominated by pion-pion annihilation and can be explained by the
decrease of hadron in-medium masses, the DLS result can not be
explained by dropping hadron masses and remains a puzzle.

\bigskip
The authors acknowledge valuable discussions with W. Cassing, C.
Gale, M. Effenberger, U. Mosel, W. Peters, M. Post, and A.
Sibirtsev throughout this study. The work of ELB was supported by
BMBF, GSI Darmstadt, while that of CMK was supported by the
National Science Foundation under Grant No. PHY-9509266 and
PHY-9870038, the Welch Foundation under Grant No. A-1358, the Texas
Advanced Research Program, and the Alexander Humboldt Foundation.

\newpage

\end{document}